\documentclass{aa}  
\usepackage{amsmath}
\usepackage{comment}
\usepackage{graphicx}
\usepackage{txfonts}
\usepackage[hidelinks,colorlinks=true,linkcolor=blue,citecolor=blue]{hyperref}
\usepackage{ulem}  
\usepackage{amsmath}

\newcommand{\be}{\begin{equation}}
\newcommand{\ee}{\end{equation}}
\newcommand{\bdm}{\begin{displaymath}}
\newcommand{\edm}{\end{displaymath}}
\newcommand{\bea}{\begin{eqnarray}}
\newcommand{\eea}{\end{eqnarray}}
\newcommand{\ba}{\begin{align}}
\newcommand{\ea}{\end{align}}

\newcommand\NB[1]{{\color{magenta}#1}}

\begin{document}
   \title{A 3+1 formalism for self-similar expanding grids in computational relativistic and Newtonian magnetohydrodynamics}
   \titlerunning{A 3+1 formalism for self-similar expanding grids}

   \author{N. Bucciantini\inst{1,2,3}\fnmsep\thanks{email: niccolo.bucciantini@inaf.it}
        \and L. Del Zanna\inst{2,1,3}
        }
        
\institute{
INAF, Osservatorio Astrofisico di Arcetri, Largo E. Fermi 5, I-50125 Firenze, Italy
\and
Dipartimento di Fisica e Astronomia, Universit\`a di Firenze, Largo E. Fermi 2, I-50125 Firenze, Italy
\and
INFN , Sezione di Firenze, Via G. Sansone 1, I-50019 Sesto Fiorentino (FI), Italy
}

   \date{Received XX XX, 20XX}

 
  \abstract
   {Astrophysical systems frequently undergo dramatic spatial expansion during their evolution. Simulating these systems with standard conservative codes for hydrodynamics and magnetohydrodynamics (MHD) using fixed grids is inherently challenging, particularly when attempting to preserve high resolution across vast changes in physical scale.}
   {While moving or expanding meshes, as well as adapting mesh refinement techniques, can address this resolution issue, they introduce computational complexity and are notoriously difficult to implement robustly—especially when preserving the solenoidal condition in magnetized systems. We aim to overcome these barriers by providing an easily implementable strategy that empowers the community to achieve high-fidelity results and establishes expanding-grid approaches as a highly accessible standard practice.} 
   {Leveraging the gauge freedom of the 3+1 formalism for MHD on curved manifolds—a foundation common to all general relativistic fluid dynamics codes—we demonstrate how an expanding coordinate system can be implemented purely through a simple modification of the underlying metric. This requires zero modifications to the numerical mesh or the core structure of the code, and the necessary addition or modification of metric terms can be easily accomplished even in Newtonian MHD.}
   {We apply this framework to standard test problems in both relativistic and Newtonian hydrodynamics and MHD, directly comparing the results of fixed and expanding coordinates. Our tests demonstrate the full consistency, accuracy, and robustness of this metric-based expansion approach, as the original conservative formulations is fully preserved.}
   {Through a simple coordinate transformation and a re-definition of the metric tensor, any general relativistic MHD code can be effortlessly transformed into an expanding-grid code. This approach elegantly guarantees both fundamental conservation properties and the solenoidal condition for the magnetic field, and can be readily extended to include non-ideal effects.}

\maketitle
\nolinenumbers

\section{Introduction}
\label{sec:intro}

Many astrophysical systems undergo large changes in size during their evolution, often traversing dynamical ranges that span several orders of magnitude. Prominent examples include explosive environments such as gamma-ray bursts \citep{Berger_13a,Granot_Horst_14a,Bromberg_Tchekhovskoy_16a}, kilonovae \citep{Metzger_17a}, supernovae \citep{Burrows_Vartanyan_21a}, and supernova remnants \citep{Orlando+16a,Bamba_Williams_22a}. Further examples arise from confined outflows, such as stellar wind bubbles \citep{Dwarkadas_23a}, pulsar wind nebulae \citep{Gaensler_Slane_06a}, and jets and their cocoons \citep{Bromberg+11a}. Simulating these environments presents a computational challenge, particularly when it is necessary to maintain high spatial resolution throughout the system's entire evolutionary history.

Several techniques have been proposed to address this issue. One option is to employ stretched grids with varying spatial resolution \citep{PLUTO07,AMRVAC18}. A typical example is the use of radial-logarithmic grids in the study of pulsar wind nebulae \citep{Del-Zanna_Amato+04a}. However, this approach suffers from two serious drawbacks. First, due to the upper limit on the choice of the global timestep, proportional to the minimum value of the grid size, the high resolution at small radii severely affects the entire evolution. Second, this method is primarily effective for spherical grids; its extension to Cartesian coordinates produces grid cells with very large aspect ratios, which are prone to generating numerical artifacts \citep{Diskin08y}. More sophisticated techniques employ adaptive mesh refinement (AMR) to dynamically adapt the grid to the expanding source \citep{Mignone2012,AMRVAC18,ATHENA20}. Implementing AMR, however, is notoriously challenging, both regarding the complex grid management and the strict requirement to conserve energy and momentum (as well as maintaining the solenoidal condition of the magnetic field in MHD \citep[e.g.][]{Balsara01a}). Alternatively, one can adopt approaches based on moving or expanding meshes, where the coordinates of the computational grid points change in time \citep{Springel10a,Marinacci+18,Habegger+21a,Mandal+23a,Heitsch+25a}. The primary difficulty with this latter method is that altering the computational mesh itself requires extreme care to ensure the conservation of physical quantities and the divergence-free nature of the magnetic field. This is typically achieved by introducing complex correction terms to the fluxes and the induction equation \citep{Heitsch+25a}. Ultimately, the common bottleneck for all of these approaches is that their implementation is mathematically and computationally intricate, typically requiring extensive modifications to the core structure of the code.

We present an approach based on a simple coordinate transformation that allows any general relativistic fluid dynamics code to operate seamlessly over an expanding grid. This framework naturally accommodates both relativistic hydrodynamics (HD) and magnetohydrodynamics (MHD), and can be trivially applied to Newtonian regimes by simply adopting a Newtonian energy-momentum tensor. In the present work the coordinate transformation is rigorously achieved by exploiting the gauge freedom of general relativity regarding the frame-dragging terms (shift vectors), combined with the gauge invariance of the Eulerian $3+1$ formalism \citep[e.g.][]{Gourgoulhon}. Consequently, this method retains all the rigorous mathematical properties of a traditional fixed-grid Eulerian cod, ensuring exact energy-momentum conservation, preserving the solenoidal condition of the magnetic field, and maintaining both numerical accuracy and extensibility. Given that general relativistic MHD codes are widely available and most of them adopt the $3+1$ framework, first rigorously introduced for the ``Eulerian Conservative High Order'' code \texttt{ECHO} \citep{ECHO07}, our approach offers a highly convenient, robust, and straightforward solution for the whole astrophysical simulation community \citep[e.g.][]{ECHO11,BHAC,iharm3D,GMUNU21,MHDuet25,ASTERX,Cupp+26a,Azizi+26a,ATHENAK}.

The paper is structured as follows: in section 2 the equations in the $3+1$ framework, as appropriate for conservative codes, are introduced; section 3 is devoted to the description of our novel method; in section 4 we validate it against standard benchmark numerical tests; conclusions are drawn in section 5. As far as notations are concerned, in the following we assume a signature $(-, +, +, +)$ for the spacetime metric and we use Latin letters $i, j, k, ...$ for 3D spatial tensor components in the $3+1$ Eulerian split, where $\gamma_{ij}$ is the spatial 3-metric (with determinant $\gamma$), the Levi-Civita pseudo-tensor is $\epsilon_{ijk}=\!\!\sqrt{\gamma}[ijk]$ ($\epsilon^{ijk}=[ijk]/\!\!\sqrt{\gamma}$), where $[123]=+1$. Moreover, we set $c = 1$ and all $\sqrt{4\pi}$ factors will be absorbed in the definition of the electromagnetic fields.

\section{Equations of $3+1$ GRMHD}

The key idea of the Eulerian 3+1 splitting is to introduce a foliation of time by space-like surfaces. If $[t,x^i]$ are the spacetime coordinates adapted to  the foliation, the line element can be written in the form
\be
\mathrm{d} s^2 = -\alpha^2 \mathrm{d} t^2+\gamma_{ij}(\mathrm{d} x^i+\beta^i \mathrm{d} t)(\mathrm{d} x^j+\beta^j \mathrm{d} t),
\ee
where the lapse function $\alpha$ and the shift vector  $\beta^i$ are free gauge functions. The equation of (ideal) General Relativistic MHD (GRMHD), expressed in the 3+1 splitting and valid on any curved manyfold, can be written as a set of fluid-like balance laws in divergence form
\be
\partial_t \vec{\mathcal{U}} + \partial_j \vec{\mathcal{F}}^j = \vec{\mathcal{S}},
\label{eq:dudt}   
\ee
complemented by a curl-like equation and a non-evolutionary  divergence-free constraint, which are the induction equation and the solenoidal condition for the magnetic field
\be
\partial_t \mathcal{B}^i + [ijk] \partial_j \mathcal{E}_k = 0, \qquad \partial_i \mathcal{B}^i=0.
\label{eq:dbdt}
\ee
In the following, we choose to use the same notations adopted in the papers describing the conservative code \texttt{ECHO} for GRMHD \citep{ECHO07, ECHO24, ECHO11}. When non-ideal effects are taken into account, such as resistivity allowing for magnetic reconnection, or mean-field dynamo mechanisms to amplify magnetic fields, an equation for the electric field, a constraint for the charge density and an expression for the current must be added to the system \citep{NBLDZ13,LDZ16,LDZNB18,Tomei20}. Here we retain the ideal MHD assumption for sake of simplicity, but the proposed method for expanding grids applies unchanged to those more general cases.\\
\\
In the first set, the conservative variables and the associated fluxes are, respectively
\be
\vec{\mathcal{U}} = \sqrt{\gamma} \left[\begin{array}{c}
D \\ S_i \\  U
\end{array}\right],~~~
\vec{\mathcal{F}}^j = \sqrt{\gamma} \left[\begin{array}{c}
 \alpha D v^j - D \beta^j  \\
 \alpha W_i^j - S_i \beta^j \\
 \alpha S^j - U \,\beta^j
\end{array}\right],
\end{equation}
where the mass current 4-vector and total momentum-energy tensor for both the fluid and the electromagnetic fields has been split into purely spatial vectors and tensors, as done for the metric, as these are the quantities measured by the so-called  Eulerian, or normal, observer associated to the foliation. Notice that in the case of the physics of black holes this observer was originally identified with the  fiducial observer FIDO \citep{Thorne82a,Thorne82b}. In particular, $D$ is the mass density, with flux $Dv^i$, $W_{ij}$ the stress tensor, $S_i$ the momentum density (coincident with the energy flux, as the original momentum-energy 4-tensor was symmetric), and $U$ the energy density. 
On the other hand, the second set gives
\be
\mathcal{B}^i = \sqrt{\gamma} B^i, \qquad \mathcal{E}_i = \alpha E_i + \epsilon_{ijk}\beta^jB^k.
\ee
where the Faraday tensor for the electromagnetic field has also been split using the $3+1$ formalism. The sources appearing on the right hand side of the divergence-type equations are
\be
\vec{\mathcal{S}} =  \sqrt{\gamma} \left[\begin{array}{c}
0 \\  
 \frac{1}{2}\alpha \,\partial_i\gamma_{jk} \, W^{jk}+\partial_i\beta^j\,S_j - U\partial_i\alpha \\ 
 \alpha K_{ij}W^{ij}-S^i\partial_i\alpha
\end{array}\right],\label{eq:source_tot}
\ee
containing derivatives of the metric terms, including the extrinsic curvature $K_{ij}$, a measure of how the $3+1$ foliation surfaces behave within the 4-dimensional spacetime, depending on the so-called Lie derivative of $\gamma_{ij}$ along the Eulerian observer \citep[e.g.][]{Gourgoulhon}. It is easy to show that it can be also expressed as
\be
\alpha K_{ij} = \tfrac{1}{2}(\gamma_{ik}\partial_j\beta^k + \gamma_{kj}\partial_i\beta^k) + \tfrac{1}{2}(\beta^k\partial_k \gamma_{ij} - \partial_t \gamma_{ij}),
\label{eq:Kij}
\ee
using just standard derivatives of metric terms. Notice that in this form of the GRMHD equations, first introduced by \cite{ECHO07}, only spatial vector and tensors appear, so that is clear that one should use the 3-metric $\gamma_{ij}$ to raise or lower indices for all quantities; moreover, only standard partial derivatives of metric terms are present in source terms. 

A derivation of the $3+1$ GRMHD equations can be found in \citet{Baum03}, where, however, a non-conservative form was adopted, so the equations are not suitable for direct numerical integration by shock-capturing codes. The so-called ``Valencia'' formulation \citep[e.g.][]{Font2008} is instead fully conservative, like the one presented here and adopted in the \texttt{ECHO} code, but components of 4D vectors are employed in fluxes and complex 4D Christoffel symbols appear in source terms, so they cannot be considered as fully compliant with a rigorous $3+1$ approach. We deem that the \texttt{ECHO} formulation \citep{ECHO07,ECHO24} is both more elegant and simpler, though fully equivalent with respect to numerical integration, as the same conservative form is adopted \citep[see also][the final discussion in Sect 7.3.3]{Rezzolla-Zanotti}. In particular, the method for expanding coordinates that will be explained below can be also implemented in codes using different formulations of the equations, but the process will be more involved (especially the computation of source terms).

In the above GRMHD system, the stress tensor, momentum density, and energy density are, respectively
\begin{eqnarray}
W_{ij}& = & \rho h \Gamma^2 v_iv_j - E_iE_j - B_iB_j + [ p + \textstyle{\frac{1}{2}}(E^2+B^2) ] \gamma_{ij},
\\
\hspace{-2mm}S_i & = & \rho h \Gamma^2 v_i + \epsilon_{ijk}E^jB^k,
\\
\hspace{-2mm}U & = & \rho h \Gamma^2 - p + \textstyle{\frac{1}{2}}(E^2+B^2).
\end{eqnarray}
whereas $D = \rho\Gamma$ appears in the mass conservation law. Here, $\rho$ and $p$ are the comoving density and pressure, respectively, $h$ is the specific enthalpy, $v^i$ is the bulk flow velocity, $\Gamma =1/\sqrt{1-v^2}$ is the Lorentz factor, $E^i$ and $B^i$, are the electric and magnetic fields, again measured by the Eulerian observer. The system must be closed by providing any equation of state (EoS) among the thermodynamic quantities, either analytical or tabulated. A common choice is the ideal gas law
\be
\epsilon = \frac{p}{{\Gamma}_{\rm ad}-1} \Rightarrow 
h = \frac{\rho + \epsilon + p}{\rho} = 1+\frac{{\Gamma}_{\rm ad}}{{\Gamma}_{\rm ad}-1}\frac{p}{\rho},
\ee
with $\epsilon$ being the internal (thermal) energy density and ${\Gamma}_{\rm ad}$ the adiabatic index, typically $4/3$ for relativistic plasmas.  Moreover, notice that the electric field is not evolved in time, given that in ideal GRMHD we are assuming a perfectly conducting plasma, so the electric field in the comoving frame must vanish. The consequence is that $E_i$ (and $\mathcal{E}_i$) is a derived quantity
\be
E_i=-\epsilon_{ijk}v^jB^k, \qquad \mathcal{E}_i =  - \epsilon_{ijk}(\alpha v^j-\beta^j)B^k,
\ee
and the ideal GRMHD system is finally closed. Finally, for numerical integration also characteristic speeds are needed. In relativistic MHD these are based on the sound and the Alfvén speeds, respectively given by 
\be
c_s^2 = {\Gamma}_{\rm ad}\frac{p}{\rho h}, \qquad c_a^2 = \frac{b^2}{\rho h + b^2},
\ee
where $b^2=B^2 - E^2=B^2/\Gamma^2 + (v_iB^i)^2$ in the ideal case. \\
\\
For non-magnetized fluids, relativistic hydrodynamics is employed, which is retrieved by simply letting $E^i=B^i=0$. Hence, the momentum-energy components contains hydrodynamic quantities alone, the Alfvén speed vanishes, there is no need to integrate in time the curl-like induction equation, since $\mathcal{E}^i=\mathcal{B}^i=0$, and the solenoidal constraint is obviously automatically verified, avoiding the use of specific numerical methods to preserve this non-evolutionary condition.

\subsection{Newtonian MHD limit}

Exactly the same $3+1$ framework may be used also for Newtonian MHD, by taking the limit for small bulk velocities and assuming that the rest mass density dominates over all other forms of energy. Reintroducing the speed of light $c$, we then impose
\be
v^2 \ll c^2, \qquad p, \, B^2 \ll \rho c^2.
\ee
When these conditions apply, the Lorentz factor approaches unity ($\Gamma \to 1$), the specific enthalpy has no thermal contribution ($h \to 1$), and the electric field is negligible compared to the magnetic one ($E^2\ll B^2$), given that the ideal MHD approximation holds unchanged, $E_i=-\epsilon_{ijk}v^jB^k$. When the rest mass energy density dominates, we have 
\be
D \to \rho, \qquad U \to \rho, \qquad S_i \to \rho v_i,
\label{eq:newtini}
\ee
the latter because the Poynting flux becomes negligible. The energy equations then reduces to the conservation of mass, and does not provide other information. We then need to subtract the two equations, so the energy equation in conservative form actually becomes
\be
\partial_t \tilde{U} + \partial_i \,[\, \!\!\sqrt{\gamma} ( \alpha \tilde{S}^j - \tilde{U}\beta^j)\,] = \!\!\sqrt{\gamma}\, [\,\alpha K_{ij}W^{ij}-S^i\partial_i\alpha \,] ,
\ee
where we have defined $\tilde{S}_i = S_i - Dv_i$ and $\tilde{U} = U - D$. In this approximation, the quantities appearing in the $3+1$ Newtonian MHD system are
\begin{eqnarray}
W_{ij} & = & \rho v_iv_j - B_iB_j + (p + \textstyle{\frac{1}{2}} B^2) \gamma_{ij},
\\
\tilde{S}_i & = & (\textstyle{\frac{1}{2}}\rho v^2 + \epsilon + p + B^2)  v_i  - (v_jB^j)B_i,
\\
\tilde{U} & = & \textstyle{\frac{1}{2}}\rho v^2 + \epsilon + \textstyle{\frac{1}{2}} B^2,
\label{eq:newtfin}
\end{eqnarray}
where we can appreciate the known MHD limits when the two terms dominated by the rest mass energy density are subtracted. Here, again we let  $\epsilon = p/({\Gamma}_{\rm ad}-1)$ for an ideal gas law (usually ${\Gamma}_{\rm ad}=5/3$ in the Newtonian regime). Finally, the two characteristic speeds are given by 
\be
c_s^2 = {\Gamma}_{\rm ad}\frac{p}{\rho}, \qquad c_a^2 = \frac{B^2}{\rho},
\ee 
which are the well know definitions in classic MHD.

\section{Self-Similar Expanding coordinates}

Here we describe the ``Self-Similar Expanding'' (SSE) coordinates method, based on our $3+1$ fully covariant formulation of the equations, which applies unchanged to both the relativistic and Newtonian systems described above. Start with a Minkowski flat spacetime, and adopt a slicing where the adapted coordinates are $[\bar{t},\bar{x}^k]$, such that the line elements reads
\be
\mathrm{d} s^2 = - \mathrm{d} \bar{t}^2 + \bar{\gamma}_{ij} \mathrm{d} \bar{x}^i \mathrm{d} \bar{x}^j,
\label{eq:metric_fix}
\ee
where the 3-metric, not necessarily diagonal, might be a function of position but not of time $\bar{\gamma}_{ij} = \bar{\gamma}_{ij}(\bar{x}^k) $. This is typical for special relativistic MHD, but it also applies in the Newtonian limit. Examples are the standard cases of Cartesian, cylindrical or spherical coordinates, and the related metrics. The gauge functions $\bar{\beta}^i$ are identically zero, so that their derivatives vanish in the source terms of the MHD equations, as well as the extrinsic curvature, $\bar{K}_{ij}=0$. Note that this condition applies not just to flat space-times, but to any static foliation, like those associated to the Schwarzschild solution in radial-spherical or isotropic coordinates.  On the other hand the spatial derivatives of $\bar{\gamma}_{ij}$ can appear in the source terms. Similarly, $\sqrt{\bar{\gamma}}$ may depend on space, but it may be taken out from time derivation of conservative variables $\mathcal{U}$. For the lapse function, Eq.~\ref{eq:metric_fix} strictly demands $\bar{\alpha}=1$, so that its spatial derivatives will also vanish. However, it is possible to retain them just in the source terms adopting the so called ``weak field regime'' of gravity, for which $\bar{\alpha}=1$ and $\nabla \bar{\alpha}\neq 0$, which is (minus) the gravity acceleration vector. Assuming $\bar{\alpha}=1$ and $\bar{\beta}^i =0$ amounts to selecting a coordinate system that is attached to the Eulerian observer, and for this reason we will refer to these coordinates, and the related metric quantities as ``fixed''.

The novel set of SSE coordinates $[t,x^k]$ are defined by the following transformation
\be
\bar{t} \equiv t, \qquad \bar{x}^i= a_i(t)\,x^i,
\label{eq:SSE}
\ee
where the function $a_i(t)$, providing the time evolution (not necessarily an expansion but possibly a contraction or even a non monotonic behavior in time), may be different for each of the three spatial coordinates ($a_i=1$ in case the corresponding coordinate is not stretched). Notice that in any expression where this separation of variables is present, the Einstein notation for repeated indexes does not apply. A typical choice for a linear expansion, ensuring synchronization of coordinates at $t=0$, is
\be
a_i(t) = 1 + t/\tau,
\label{eq:tau}
\ee
for any or all coordinates, where $\tau$ is the timescale of grid expansion. The transformations in Eqs.~\ref{eq:SSE}-\ref{eq:tau} were first adopted in the $3+1$ formulation of the "Expanding Box Model'' (EBM) for the solar wind \citep{ASTRONUM12,LDZ15}, where the rate of expansion was named $\epsilon=1/\tau$. Thanks to the above change of coordinates, covariant and contravariant components of spatial vectors and tensors transform as
\be
V_i = a_i \bar{V}_i, \qquad V^i = (1/a_i) \,\bar{V}^i,
\ee
with $V^i$ a generic spatial vector. 
The fundamental step in the SSE approach is the introduction of shift vectors in the new coordinate system, even if absent in the initial Minkowski metric.  Differentiation of Eq.~\ref{eq:SSE} leads to
\be
\mathrm{d} \bar{x}^i = a_i \,(\mathrm{d} x^i + \beta^i \mathrm{d} t),
\ee
so the line elements in the SSE coordinates, in the canonical $3+1$ form, becomes
\be
\mathrm{d} s^2 = - \mathrm{d} t^2+\gamma_{ij}(\mathrm{d} x^i+\beta^i \mathrm{d} t)(\mathrm{d} x^j+\beta^j \mathrm{d} t),
\ee
where
\be
\alpha = 1, \qquad
\beta^i = (\dot{a_i}/a_i) x^i, \qquad
\gamma_{ij} = a_ia_j\,\bar{\gamma}_{ij}(a_k\,x^k), 
\ee
and also
\be
\sqrt{\gamma} = a_1a_2a_3 \sqrt{\bar{\gamma}}, 
\ee
where all quantities are now function of the SSE coordinates $[t,x^k]$ and where a dot indicates time derivation ($\dot{a}_i=\partial_t a_i$). Clearly $\beta^i=0$ if the corresponding coordinate $x^i=\bar{x}^i$ is not stretched in time. Moreover if the original fixed coordinate system was orthogonal, the new SSE coordinate system will remain orthogonal. 
 
 When using these coordinates in the GRMHD equations, one should also compute spatial derivatives of the metric terms. Thanks to the previous relations we have
 \be
 \partial_k\alpha =0, \qquad \partial_k \beta^i = (\dot{a_i}/a_i)\gamma_k^{\,i}, \qquad \partial_k \gamma_{ij} = a_i\,a_j\,a_k \frac{\partial \bar{\gamma}_{ij}}{\partial \bar{x}^k},
 \ee
 with spatial derivatives of the metric tensor vanishing only for a choice of initial, non-stretched,  Cartesian coordinates $\bar{x}^i$. In addition, notice that
 \be
 \partial_t \gamma_{ij} = \left(\frac{\dot{a_i}}{a_i} + \frac{\dot{a_j}}{a_j}\right)\gamma_{ij} + a_ia_j\,a_k\,\beta^k \frac{\partial \bar{\gamma}_{ij}}{\partial \bar{x}^k},
 \ee
 and it is easy to check from Eq.~\ref{eq:Kij} that $K_{ij}=0$ in the SSE-coordinate system too. 
 The source terms for the divergence part of GRMHD equations then simplify to
\be
\vec{\mathcal{S}} =  \sqrt{\gamma} \left[\begin{array}{c}
0 \\  
 \frac{1}{2} \partial_i\gamma_{jk} \, W^{jk} + (\dot{a_i}/a_i)\,S_i  \\ 
 0
\end{array}\right],\label{eq:source_sse}
\ee
in which only a new term proportional to $S_i$ appears, in the momentum equation, with respect to equations in the fixed coordinate system $[\bar{t},\bar{x}^k]$.

Notice that the shift vectors $\beta^i$ were not implemented explicitly in the (Newtonian) metric for EBM, where only the $\gamma_{ij}$ terms were affected by the presence of the expansion factors $a_i(t)$, so that a term $K_{ij} \propto \partial_t \gamma_{ij}$ appeared in the energy equation.

   \begin{figure*}[t]
        \centering
        \includegraphics[width=6cm]{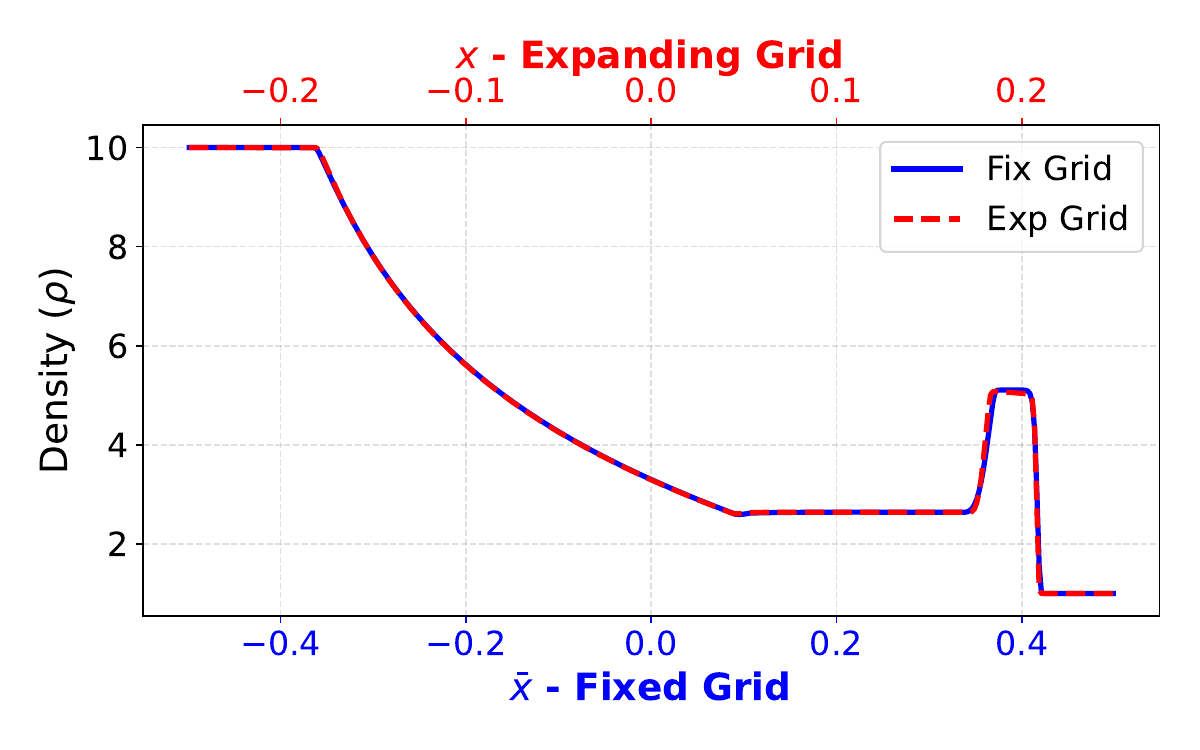}\includegraphics[width=6cm]{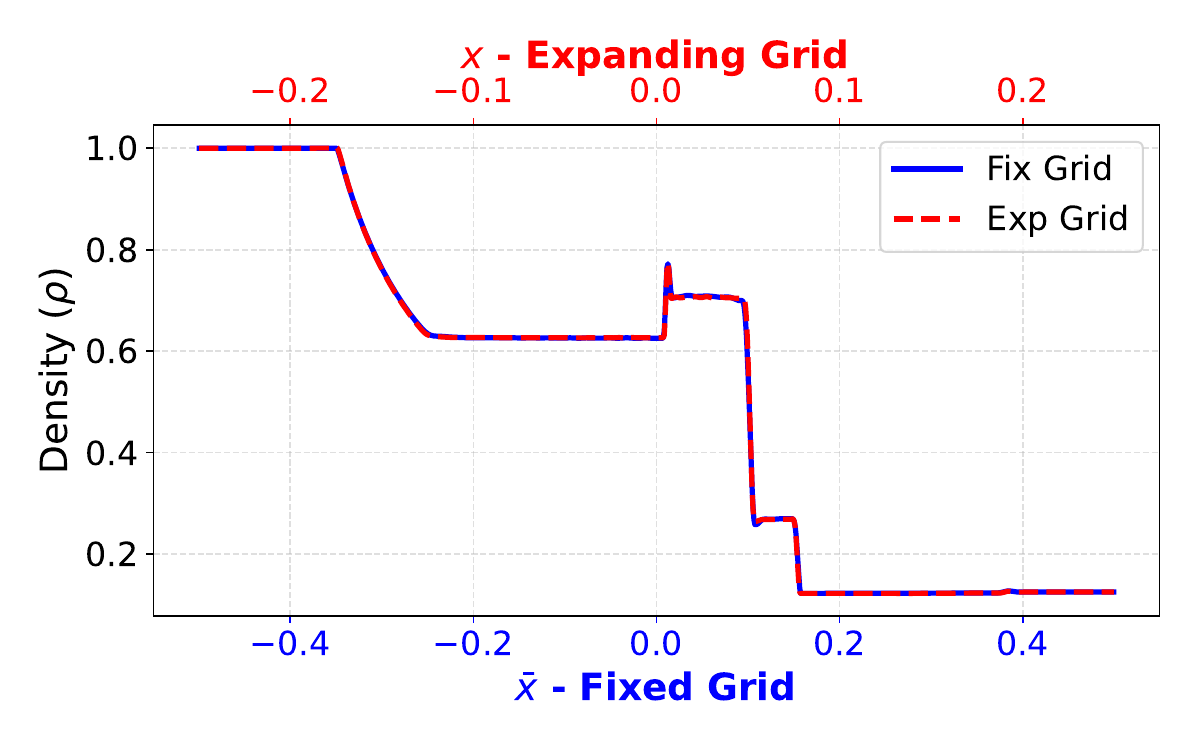}
        \includegraphics[width=6cm]{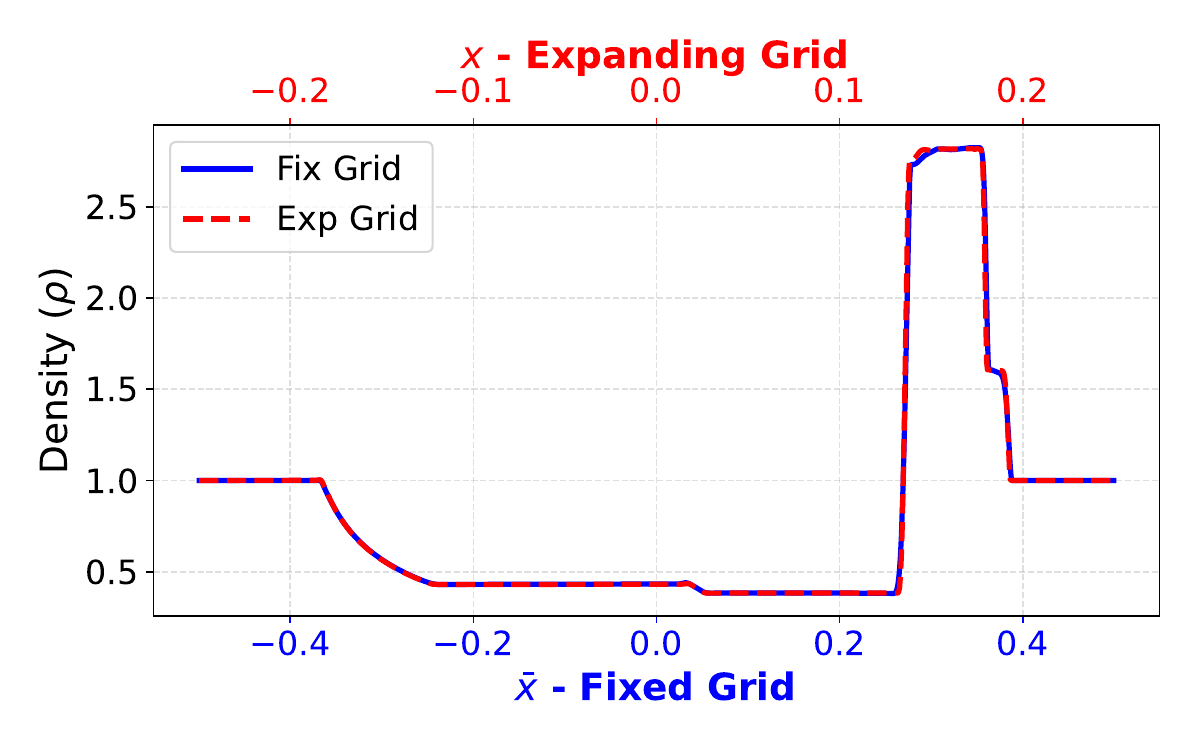}
        {\caption{1D cartesian shock-tube problems. Comparison of the comoving density $\rho$ between fixed-coordinates (solid blue line) vs SSE coordinates (dashed red line). Left panel: Sod problem at $t=\tau=0.5$, over 400 grid zones for both the fixed and expanding grid.  Central panel: Brio \& Wu problem at $t=\tau=0.4$, over 1000 grid zones for both the fixed and expanding grid.  Right panel: blast wave problem at $t=\tau=0.4$, over 1000 grid zones for both the fixed and expanding grid.}
       \label{fig:1}}
    \end{figure*}

\section{Numerical validation of SSE grids}
\label{sec:test}

We present a suite of 1D and 2D benchmark problems to demonstrate the performance and efficacy of our approach. All tests were performed using the \texttt{ECHO} code for GRMHD, assuming ideal conditions and a standard $\Gamma$-law ideal gas equation of state. Given that the typical astrophysical problems in which we are interested are computed using standard orthogonal coordinates (Cartesian, cylindrical, or spherical), we restrict our analysis to orthogonal coordinate systems and adopt simple linear expansion laws $a(t)-1 \propto t$ for the stretched SSE grids, unless stated otherwise. All problems are special relativistic ones, where a Minkowski flat spacetime is assumed for the non-expanding metric. Notice that we only select expanding factors where $a(0)=1$, so that initial conditions, including vector components, are identical in fixed grids $\bar{x}^i$ and SSE ones $x^i$.

All tests were executed using the Harten, Lax \& van Leer (HLL) Riemann solver \citep{HLL}, featuring second-order discretization in time and space and employing a Monotonized Central (MC) limiter for spatial reconstruction \citep{van_Leer_77a}. The Courant–Friedrichs–Lewy (CFL) parameter is set to $0.3$. Given that the robustness, accuracy and overall performances of the \texttt{ECHO} code have been tested elsewhere, we limit this section to few test cases illustrative of the SSE approach.

\subsection{1D Cartesian shock tubes}

For the simple one-dimensional Cartesian shock-tube tests, we adopt the following fixed-grid and SSE-grid metrics
\bea
& \mathrm{d} s^2 \!\!\! & = - \mathrm{d} \bar{t}^2 +\mathrm{d} \bar{x}^2 +\mathrm{d} \bar{y}^2 +\mathrm{d} \bar{z}^2 \nonumber \\
& & = - \mathrm{d}t^2 + (1+t/\tau)^2 \left(\mathrm{d}x + \frac{x}{1+t/\tau}\,\frac{\mathrm{d} t}{\tau} \right)^2 + \mathrm{d}y^2 + \mathrm{d}z^2,
\eea
where the coordinate transformation along $x$ ($y$ and $z$ are not stretched) corresponds to the linear expansion law 
\be
\bar{x}=a(t)x, \qquad a(t) = 1+t/\tau,
\ee
ensuring that the fixed and expanding coordinate systems are strictly synchronized at $t=0$. The specified value of $\tau$ determines the relative scaling between the two coordinate systems at the conclusion of the numerical run. In all tests we map the spatial domains as
\be
\bar{x} \in [-0.5,0.5], \qquad x \in [-0.25,0.25]
\ee
so the expansion is of a factor 2, and if one wants $a(t)=2$ by the end of the simulation, the timescale $\tau$ must be taken equal to the final time $t$ (values may differ from one test to another, see below). In addition, all shock tubes have initial static conditions, so $v^x=v^y=v^z=0$.

In Fig.~\ref{fig:1} we show the results of three standard shock tube problems in relativistic HD and MHD \citep{ECHO02,ECHO03}. A HD Sod test ($\Gamma_{\rm ad}=5/3$, $t=\tau=0.5$), with
\begin{equation}
    \begin{cases}
          \quad p = 13.3,\, \quad \rho = 10.0, \quad{\rm for} \quad x<0\\
         \quad p = 10^{-6}, \quad \rho = 1.0,\,\,\, \quad{\rm for} \quad x>0
    \end{cases}
\end{equation}
The \citet{Brio_Wu88a} shock tube MHD problem ($\Gamma_{\rm ad}=2$, $t=\tau=0.4$), with $B^x=0.5$, $B^y=0.5$, and
\begin{equation}
    \begin{cases}
          \quad p = 1.0, \quad \rho = 1.000, \quad B^z=\,\,\,\,1.0, \quad{\rm for} \quad x<0\\
         \quad p = 0.1, \quad \rho = 0.125,  \quad B^z=-1.0, \quad{\rm for} \quad x>0
    \end{cases}
\end{equation}
A 1D blast wave test ($\Gamma_{\rm ad}=5/3$, $t=\tau=0.4$), with $\rho =1$, $B^x=0.5$,  and
\begin{equation}
    \begin{cases}
          \quad p = 1.0, \quad B^y =B^z = 6.0, \quad{\rm for} \quad x<0\\
         \quad p = 0.1, \quad B^y =B^z = 0.7, \quad{\rm for} \quad x>0
    \end{cases}
\end{equation}
Across all considered test cases, we observe excellent quantitative agreement between the fixed and expanding coordinate frameworks. Notably, because the expanding-grid approach intrinsically maintains a higher average effective spatial resolution as the physical system dynamically evolves, these simulations resolve fine-scale structures with somewhat greater sharpness than their corresponding fixed-grid counterparts. At the computational domain boundaries, we apply a zeroth-order extrapolation scheme to the fluid variables, whereas the metric tensor is explicitly extended utilizing its exact analytical formulation. Given that our SSE coordinate transformation does not change the Eulerian observer, the boundary conditions on the fluid variables (density, pressure, orthonormal components of velocity and magnetic field) are set according to the Eulerian observer (i.e.  the extrapolation procedures apply directly to quantities relative to that observer) at the location of the moving boundary.

   \begin{figure}[b]
        \centering
        \includegraphics[width=8.5cm]{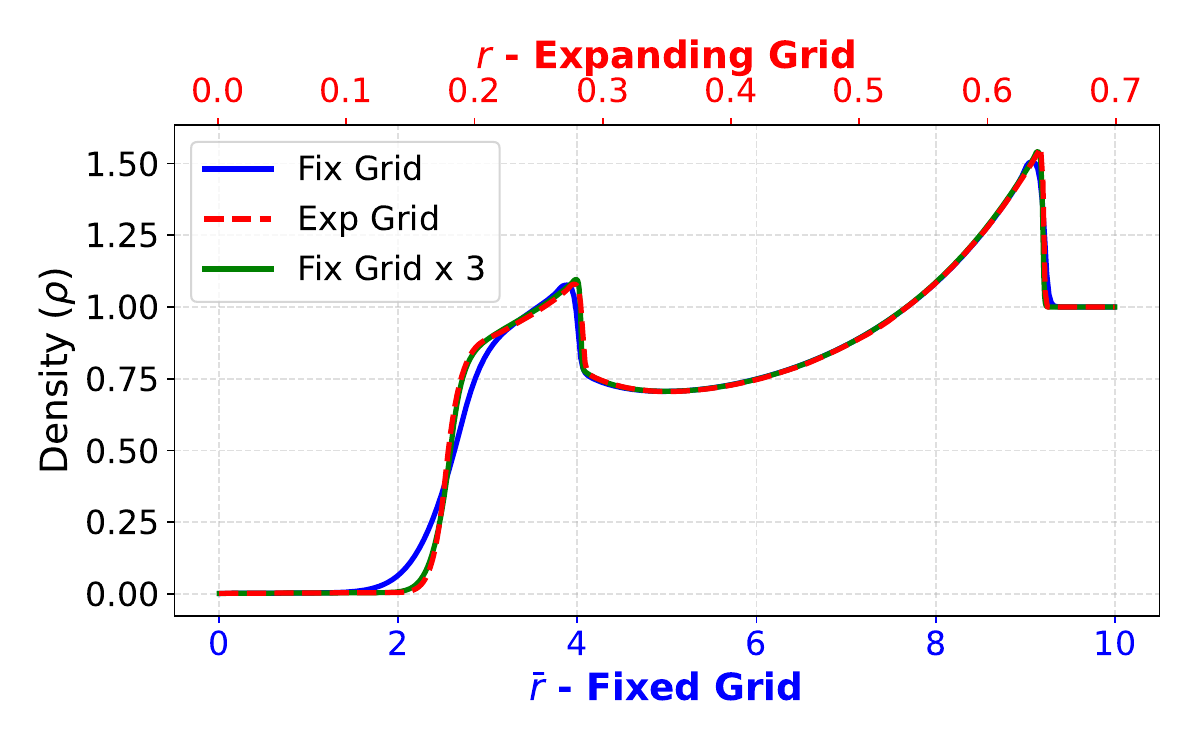}
        {\caption{1D spherical blast wave problem. Comparison of the comoving density $\rho$ at $t=12$ between fixed-coordinates (solid blue line) versus SSE coordinates (dashed red line) using $\tau=0.55$ and $\kappa=0.85$, over 400 zones for both grids. The green line is a solution in fixed coordinates with an enhanced resolution, using 1200 grid points. }
       \label{fig:2}}
    \end{figure}

\subsection{1D spherical blast wave}

We next consider a spherically symmetric HD relativistic blast wave \citep{ECHO02}. This test serves to illustrate the framework's robustness when handling both non-Cartesian coordinate systems and non-linear time-dependent expansion laws. The example is just 1D, but many 3D applications in spherical coordinates are expected to require a stretching of the radial coordinate alone, so this test is representative of a wider class of situations where radial expansion is encountered. The line elements for the fixed and SSE metrics are defined as
\bea
& \mathrm{d} s^2 \!\!\! & = - \mathrm{d} \bar{t}^2 +\mathrm{d} \bar{r}^2 + \bar{r}^2 \mathrm{d} \bar{\Omega}^2 \nonumber \\
& & = - \mathrm{d}t^2 + (1+t/\tau)^{2\kappa} \left[\left(\mathrm{d}r + \frac{\kappa \,r}{1+t/\tau}\,\frac{\mathrm{d} t}{\tau} \right)^2 + r^2 \mathrm{d} \Omega^2 \right],
\eea
where the fixed-grid coordinates are $[\bar{t}, \bar{r}, \bar{\theta}, \bar{\phi}]$,  and $[t, r, \theta, \phi]$ the SSE ones, whereas $\mathrm{d} \bar{\Omega}$ and $\mathrm{d}\Omega$ are the usual solid angle elements in the two system of coordinates. The above transformation refers to an expansion factor, for the radial coordinate alone
\be
\bar{r} = a(t) r, \qquad a(t)=(1+t/\tau)^\kappa
\ee
with $\kappa=0.85$ a coefficient introduced to provide a milder (here $\kappa <1$) expansion rate. Notice that the angular coordinates $\theta$ and $\phi$ are not stretched, so also $\mathrm{d}\Omega = \mathrm{d} \bar{\Omega}$. At the initial time $t=0$, the fluid is at rest ($v^x=v^y=v^z=0$) with a uniform density $\rho=1.0$. The adiabatic coefficient is $\Gamma_{\rm ad}=4/3$. The initial pressure profile represents a highly overpressurized core
\begin{equation}
    p = \begin{cases}
          1000, & \text{for } r < 0.4 \\
          1.0,  & \text{for } r > 0.4
    \end{cases}
\end{equation}
The evolution is followed up to a final time $t=12$, at which point the blast wave has expanded to $\bar{r}\simeq 9$, representing an increase in physical size by a factor of more than 20 compared to the initial configuration. To contain the whole evolution, the grids are
\be
\bar{r} \in [0,10], \qquad r \in [0,0.7],
\ee
and the expansion timescale is chosen as $\tau=0.55$. In Fig.~\ref{fig:2}, we compare the results obtained using fixed coordinates against those obtained employing SSE coordinates. Crucially, in the expanding framework, the shock front traverses a relative coordinate distance that increases by a much smaller factor. Using $\kappa=0.85$ and choosing a timescale $\tau=0.55$, this factor is only $\simeq 1.6$. This stark contrast highlights the profound advantage of the expanding-grid approach: it naturally maintains a high, constant relative resolution across the entire dynamic evolution of the system. We estimate that to achieve a comparable level of fidelity over this 20-fold physical expansion, a fixed-grid simulation would require at least three times the number of computational zones (see the green line in the figure). 

Of course, assessing an equivalent fixed grid resolution might be problem and feature dependent. Features like contact discontinuities, that form at early times, are the ones that benefit more from an expanding grid that ensures higher resolution since earlier times. Others, especially shocks, that are reformed continuously are more sensitive to the final resolution. 
Consequently, our method not only captures the sharp features of the blast wave with superior clarity, but achieves this high-resolution accuracy at a fraction of the computational cost.  In relativistic simulations the speeds entering the CFL condition are always $\sim c$, and the time step effectively scales with the cell size $\Delta l$ as $\Delta t \propto 1/ \Delta l$ (in 1D problems).  Assuming that the code performs the same number of operations to update a computational cell for fixed and SSE grids, for a linear or quasi-linear expansion, the computational times $T$ for fixed and SSE simulation should scale as
\begin{eqnarray}
    \frac{T_{\rm sse} }{ T_{\rm fix} }\approx \frac{N_{\rm sse} }{ N_{\rm fix}}  \frac{ \Delta l_{\rm fix} } {\Delta l_{\rm sse} }\bigg|_{\rm final} \times \ln[\eta_{\rm exp}], \label{eq:scal}
\end{eqnarray}
where $N$ is the numbers of grid zones, the grids are uniformly spaced, the cell sizes are compared at the final times and $\eta_{\rm exp}$ quantifies how much the SSE gid has expanded and accounts for the fact that during the simulation $\Delta l_{\rm sse} $ increases. In this specific test, comparing the SSE simulation with the one on a fixed grid with three times higher number of points, Eq.~\ref{eq:scal} gives a relative ratio $\sim 0.3$, which agrees with the results of the \texttt{ECHO}  runs.

While this may look of little benefit in simple 1D radial applications, it has a crucial impact on the amount of resources needed in 3D simulations,  or for large expansions, where SSE grids allow reducing substantially the number of grid zones.

   \begin{figure*}
        \centering
        \includegraphics[width=9cm, bb =10 75 515 500, clip]{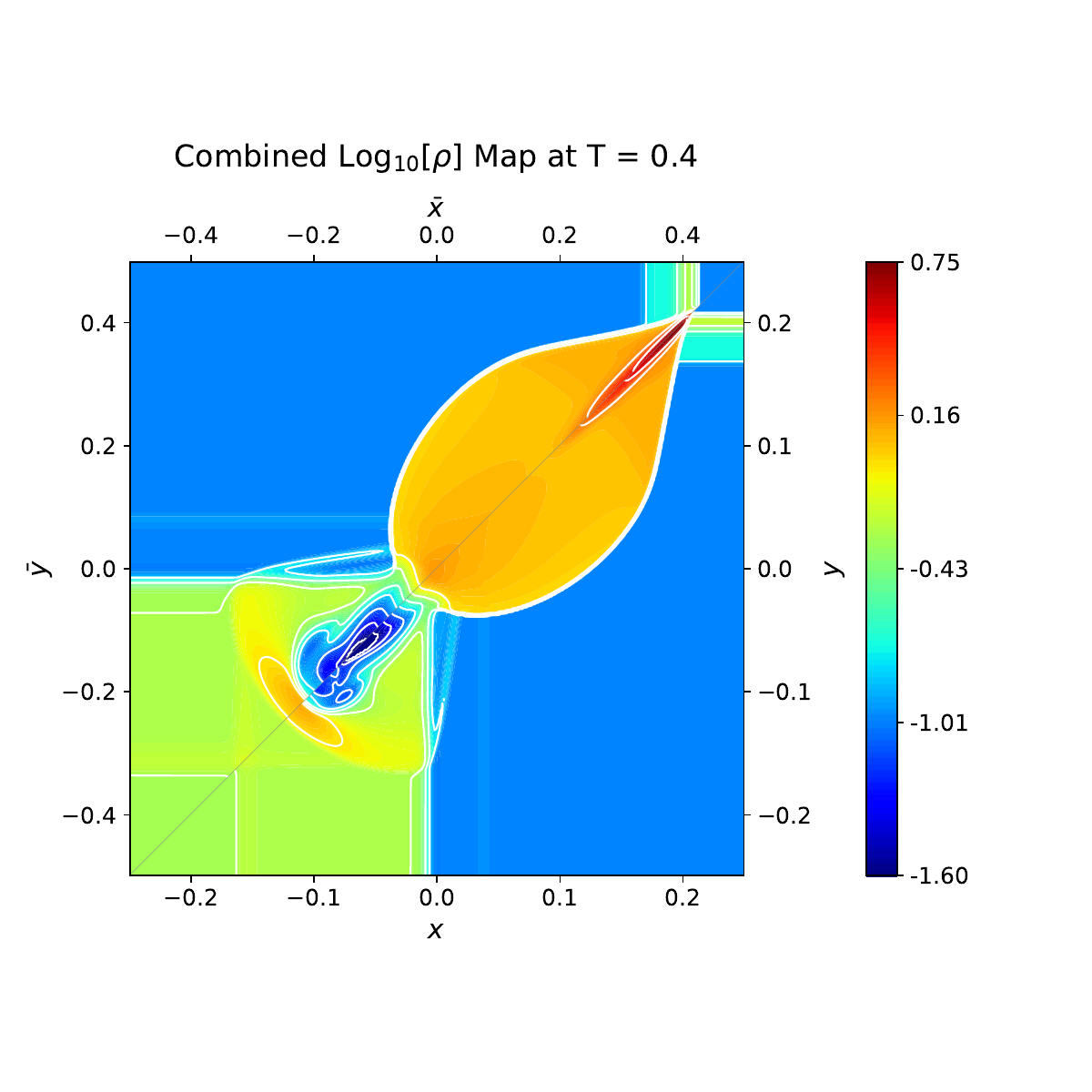}
        \includegraphics[width=9cm, bb =10
        75 515 530, clip]{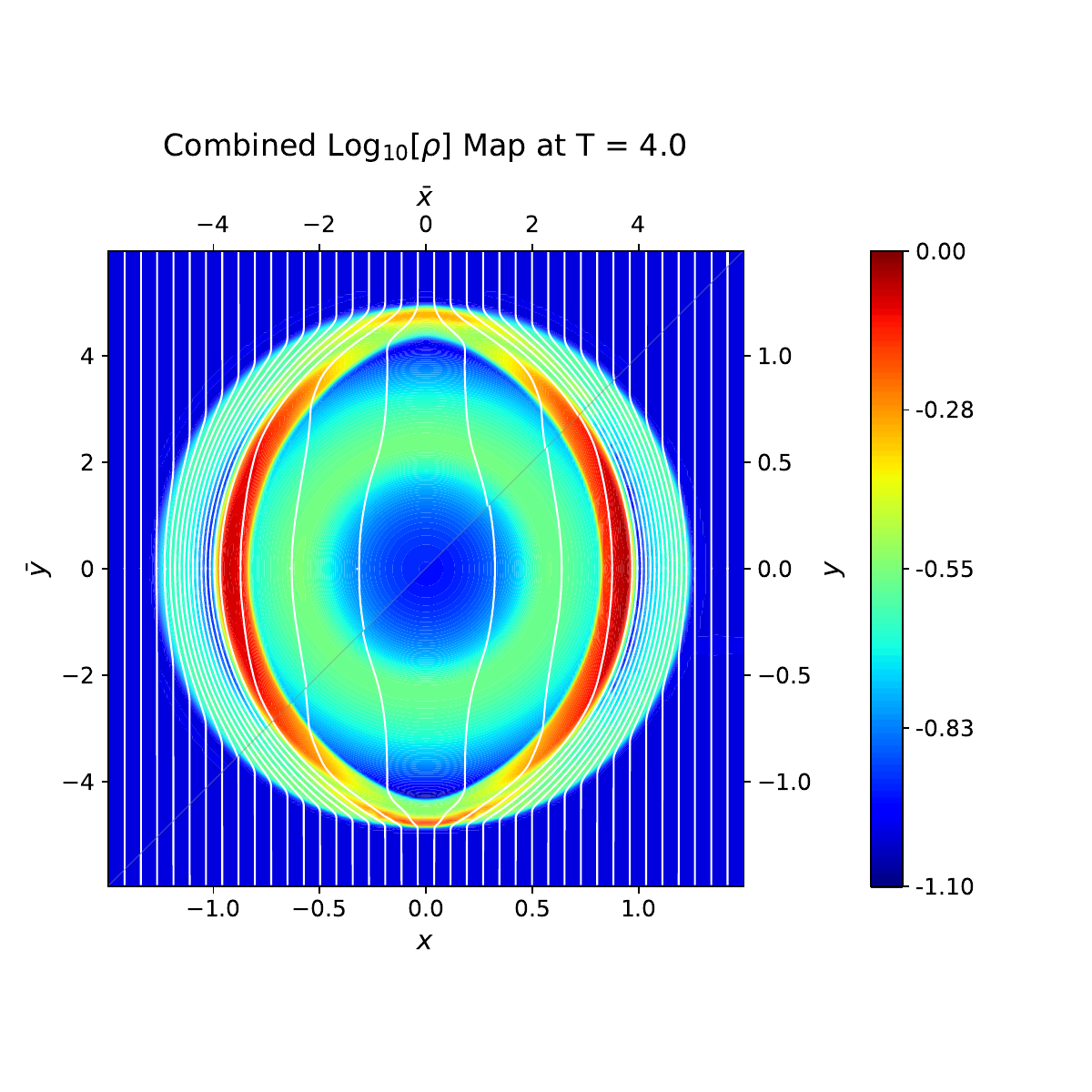}
        {\caption{Comparison of results for 2D Cartesian problems using SSE and fixed coordinate systems. The lower-right half area (below the  counter-diagonal) shows the simulation results in the SSE coordinates $x-y$, the upper-left half area (above the  counter-diagonal) shows the analogous outcome for the fixed coordinates $\bar{x}-\bar{y}$. The color map represents the logarithm of the commoving density. Left panel: 2D Riemann problem. Contours represent density iso-levels.  Right panel: 2D blast wave. Contours indicate density iso-levels, white thin lines represent magnetic field lines. Both simulations were performed on a $400 \times 400$ grid in both the fixed and SSE-coordinates.}
       \label{fig:3}}
    \end{figure*}
    
\subsection{2D Cartesian problems}

For simple Cartesian 2D shock-tube tests we adopt the same stretching already employs for 1D Cartesian tests, now imposed on two coordinates. The line elements for fixed-grid and expanding-grid are then
\bea
& \mathrm{d} s^2 \!\!\! & = - \mathrm{d} \bar{t}^2 +\mathrm{d} \bar{x}^2 +\mathrm{d} \bar{y}^2 +\mathrm{d} \bar{z}^2 \nonumber \\
& & = - \mathrm{d}t^2 + \mathrm{d}z^2 +  \\
& & (1+t/\tau)^2\left[
\left(\mathrm{d}x + \frac{x}{1+t/\tau}\,\frac{\mathrm{d} t}{\tau} \right)^2 + \left(\mathrm{d}y + \frac{y}{1+t/\tau}\,\frac{\mathrm{d} t}{\tau} \right)^2 \right] , \nonumber
\eea
where the same coordinate transformation applies to both $x$ and $y$ ($z$ is not stretched), and the usual linear expansion law 
\be
\bar{x}=a(t)x,\quad \bar{y}=a(t)y, \qquad a(t) = 1+t/\tau
\ee
is chosen. As usual, the value of $\tau$ depends on the expansion factor that we want to achieve at the final time $t$ of the simulation.

The first test that we consider here is a 2D relativistic HD Riemann problem \citep{ECHO02}, where four different uniform states are initialized and then the system is free to evolve. Initial conditions at $t=0$ for this test are
\begin{eqnarray}
    \begin{cases}
        \quad(p,v^x,v^y,p)^\mathrm{NE} &= (0.1,0,0,0.01)\nonumber\\
        \quad(p,v^x,v^y,p)^\mathrm{NW} &=(0.1,0.99,0,0.1)\nonumber\\
        \quad(p,v^x,v^y,p)^\mathrm{SW} &=(0.5,0,0,0.1)\nonumber\\
        \quad(p,v^x,v^y,p)^\mathrm{SE} &=(0.1,0,0.99,1)\nonumber\\
    \end{cases}
\end{eqnarray}
this time with non-vanishing velocities in two quadrants, with $v^z=0$ and $\Gamma_{\rm ad}=5/3$. The position of the four states in the 2D grid corresponds to the indicated directions, and we assume stretching grids expanding of a factor $a=2$ by the end of the simulation. The grids assumed for this test are
\be
\bar{x}, \bar{y} \in [-0.5,0.5], \qquad x,y \in [-0.25,0.25],
\ee
and $\tau=t=0.4$. The results, shown in the left panel of Fig.~\ref{fig:3}, are in full agreement between those obtained using the SSE-grid and those using the fixed grid. The position of the various shocks and contact discontinuities are correctly recovered. The effect of the higher average resolution associated with the SSE-coordinate grid can be seen by looking at the position of the leading edge of the compression front at $\bar{x}=\bar{y}=0.4\rightarrow x=y=0.2$ which appears to be slightly sharped an a bit backward. Similarly we can see that the jetlet that forms around $\bar{x}=\bar{y}=-0.2\rightarrow x=y=-0.1$ is more collimated in the SSE grid, and similarly the bow-waves that it drives toward the SE quadrant is a bit thinner. Indeed, while the maximum Lorentz factor achieved on the fixed grid at the core of the jetlet is $\sim 30$, in the SSE grid it reaches $\sim 40$.

In Fig.~\ref{fig:3}, right panel, we show the results of a (rescaled) 2D relativistic MHD blast wave test \citep{ECHO07}. Here the grids are
\be
\bar{x}, \bar{y} \in [-6.0,6.0], \qquad x,y \in [-1.5,1.5],
\ee
for an output time $t=4$ and expansion timescale $\tau=1$, providing a scale factor $a=4$ by the end of the simulation. At the initial time $t=0$ the fluid is at rest ($v^x=v^y=v^z=0$), threaded by a vertical uniform magnetic field $B^x=B^z=0$, $B_yB^y = 10$, with $\Gamma_{\rm ad}=4/3$. The plasma presents a denser highly overpressurized core, with respect to the ambient medium, initialized as
\be
    \begin{cases}
          \quad p = 1000., \quad \rho = 10., &\quad{\rm for} \quad r<1\\
         \quad p = 0.5, \quad \,\,\,\,\,\rho = 0.1, &\quad{\rm for} \quad r>1
    \end{cases}
\ee
with $r^2=x^2+y^2$. Again, we see agreement in the location and amplitudes of the various structures between the fixed and SSE grids. There is some hint of a slightly sharper front along the $\bar{x}=x=0$ direction. Note, moreover, that the magnetic field lines in the two coordinate systems match, indicating that the solenoidal condition is preserved in the presence of strong shocks also when using SSE coordinates. 

   \begin{figure*}
        \centering
        \includegraphics[width=6cm, bb =5 55 428 700, clip]{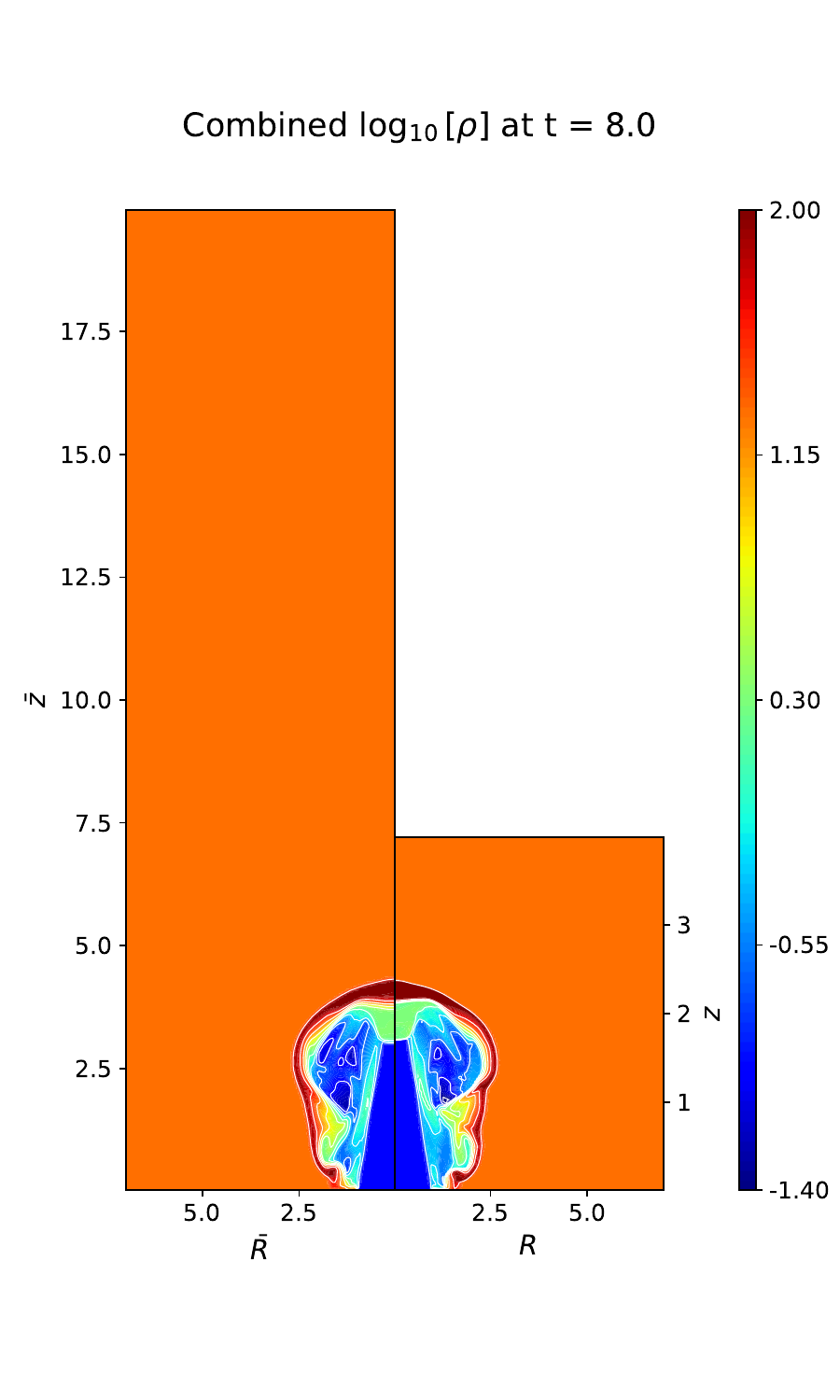}
        \includegraphics[width=6cm, bb =5 55 428 700, clip]{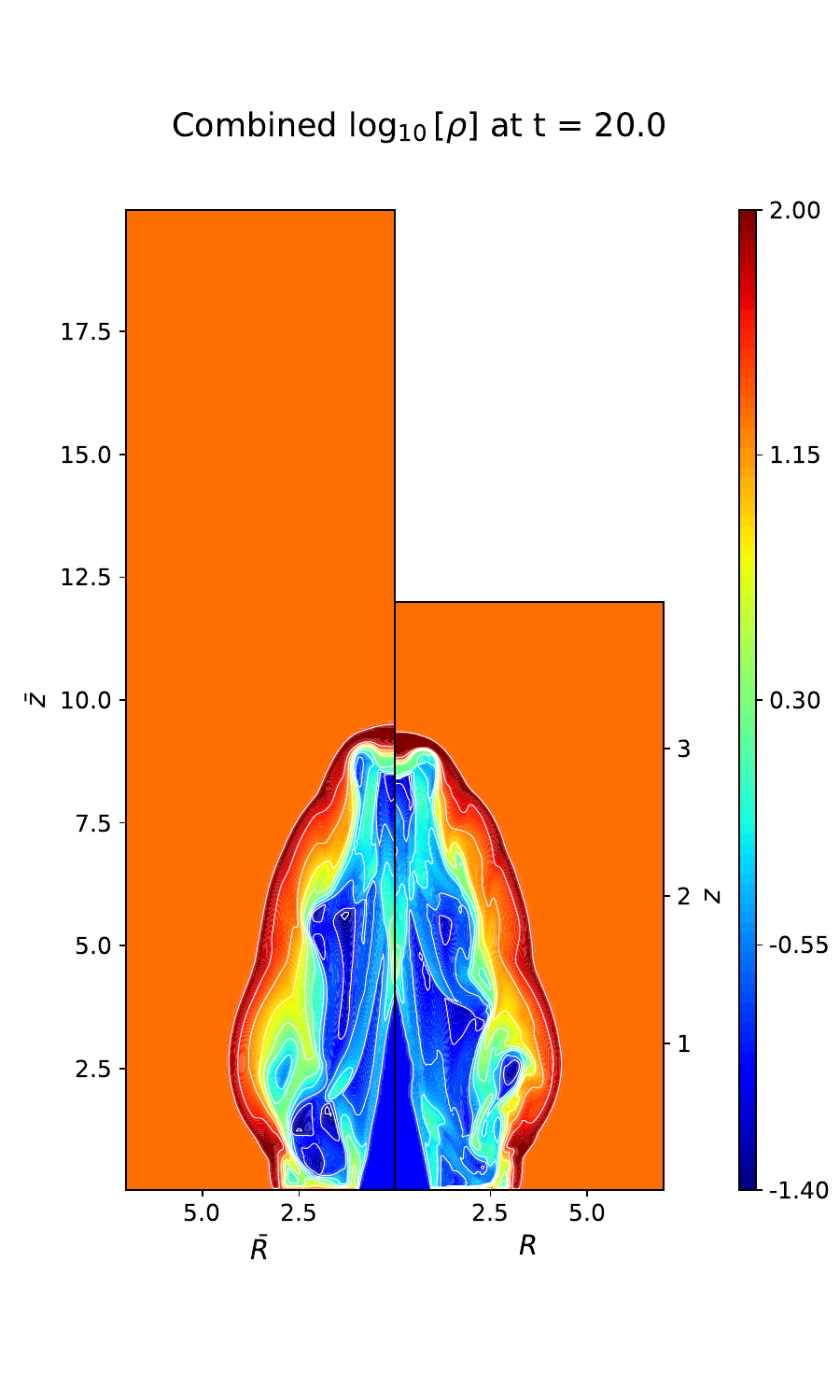}
        \includegraphics[width=6cm, bb =5 55 428 700, clip]{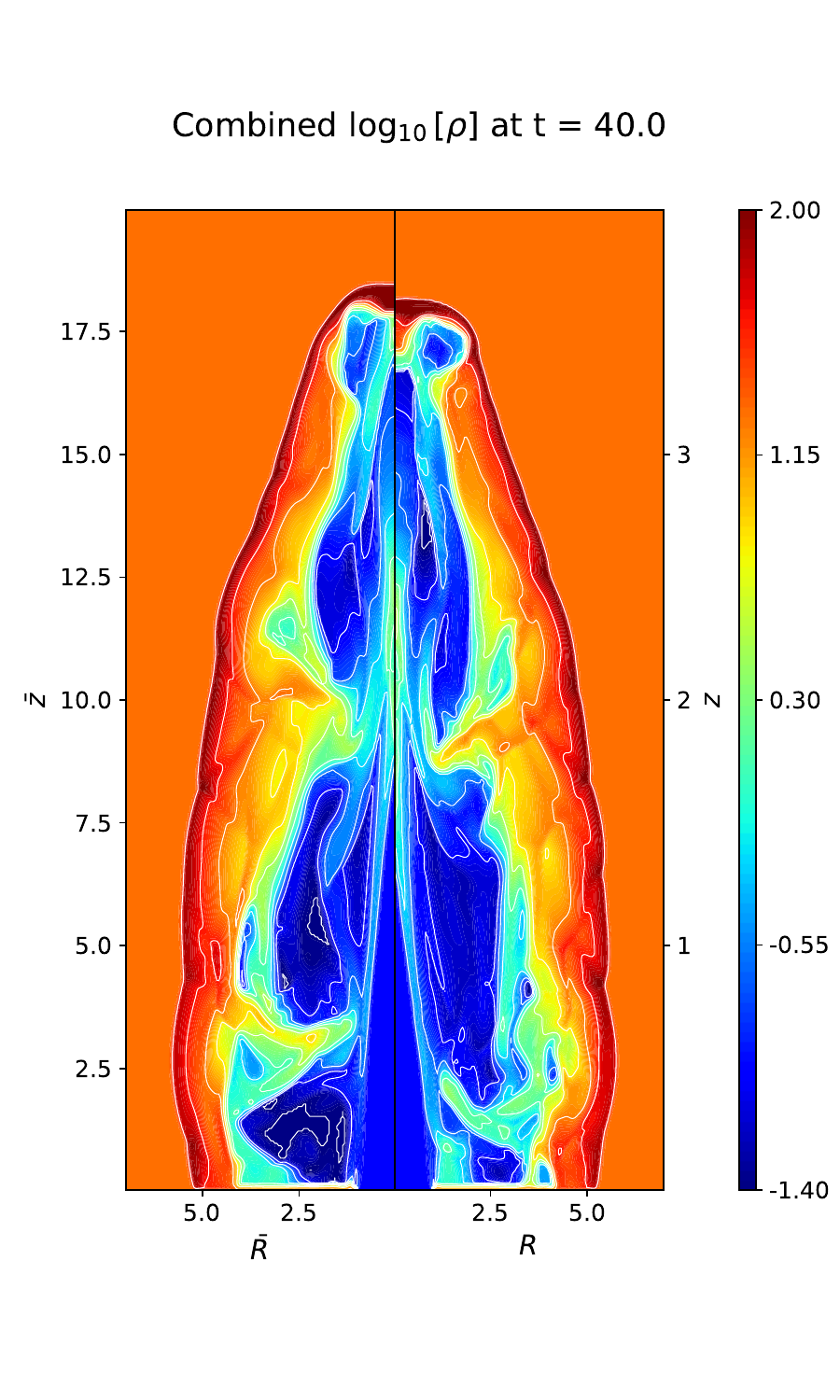}
        {\caption{2D cylindrical jet. The right half areas show the simulation results at $t=8, 20, 40$ in the SSE coordinates, the left half areas  show the analogous outcomes for the fixed coordinates. The color map represents the logarithm of the comoving density, with contours indicating density iso-levels. Simulations were performed on a grid with $140$ cells in the $\bar{R} = R$ direction and  $400$ in both the fixed $\bar{z}$ and expanding $z$ coordinates. }
       \label{fig:4}}
    \end{figure*}

\subsection{2D cylindrical jet}

As an example of non-isotropic expansion we consider the case of the propagation of an axisymmetric (unmagnetized) jet in cylindrical coordinates.  The corresponding line elements for the fixed and SSE coordinates are
\bea
& \mathrm{d} s^2 \!\!\! & = - \mathrm{d} \bar{t}^2 +\mathrm{d} \bar{R}^2 + \bar{R}^2\mathrm{d} \bar{\phi}^2 +\mathrm{d} \bar{z}^2  \\
& & = - \mathrm{d}t^2 + \mathrm{d}R^2 + R^2\mathrm{d}\phi^2 + (1+t/\tau)^2 \left(\mathrm{d}z + \frac{z}{1+t/\tau}\,\frac{\mathrm{d} t}{\tau} \right)^2, \nonumber
\eea
where the coordinate transformation along $z$ ($R$ and $\phi$ are not stretched) corresponds to the linear expansion law 
\be
\bar{z}=a(t)z, \qquad a(t) = 1+t/\tau,
\ee
with $[\bar{t}, \bar{R}, \bar{\phi}, \bar{z}]$ the standard cylindrical coordinates of the fixed grid, and $[t, R, \phi, z]$ the SSE coordinates. The domain is 
\be
\bar{R} \in [0,7],\,\bar{z} \in [0, 20], \qquad R\in [0,7],\, z \in[0,4], 
\ee
with clearly a maximum expansion factor of $a(t)=5$ at the output time $t=40$, so we assume $\tau=10$ for the present test. Reflective boundary conditions are imposed on the axis $R = 0$ and simple zero-th order extrapolation at the other boundaries, except at $z = 0$, 
where initial values are kept constant. At $t =0$ a relativistic HD jet with $v^z = 0.99$ and density 100 times less than the external static medium (with the same pressure) is initialized as follows
\bea
    \begin{cases}
\quad(\rho, v^R, v^z, p) &= (0.1, 0.99, 0.0, 0.01),\quad R < 1,z < 1\nonumber\\
\quad(\rho, v^R, v^z, p) &= (10, \,\,\,\,0.0, 0.0, 0.01),\quad {\rm outside}\nonumber
    \end{cases}
\eea

In Fig.~\ref{fig:4} we compare results using the fixed and the SSE grids, with the same number of cells along both directions: $140$ in $\bar{R} = R$ and $400$ in both the fixed $\bar{z}$ and expanding $z$ coordinates. Here we show snapshots of the evolution at three different times, $t=8, 20, 40$, to better appreciate the power of expanding grids. Clearly the numerical domain in the SSE grid $R-z$ is more suited to cover the jet evolution in time, with a larger number of computational zones inside the evolving structure, while in the fixed grid $\bar{R}-\bar{z}$ a broad portion of the domain just contains the ambient medium for large part of the simulation. At the final output, for $t=40$, when the two results can be directly compared, the overall structure, the amount of lateral expansion, the location of the head, and the position of internal oblique shocks are all in agreement. Differences can be noted in those structures that depend on the development of turbulence, and that are more sensitive to the effective resolution. In particular, the forward shock in the jet head appears to be slightly retarded when using the SSE coordinates, and this is due to a reduced diffusion at very early times when the difference in effective resolution is stronger. Note that while with the fixed grid the jet expands by about a factor 18, with the SSE grid the expansion is just a factor 3.5.

\subsection{Newtonian 2D MHD blast wave}

\begin{figure}
        \centering
        \includegraphics[width=9cm, bb =10 75 515 530, clip]{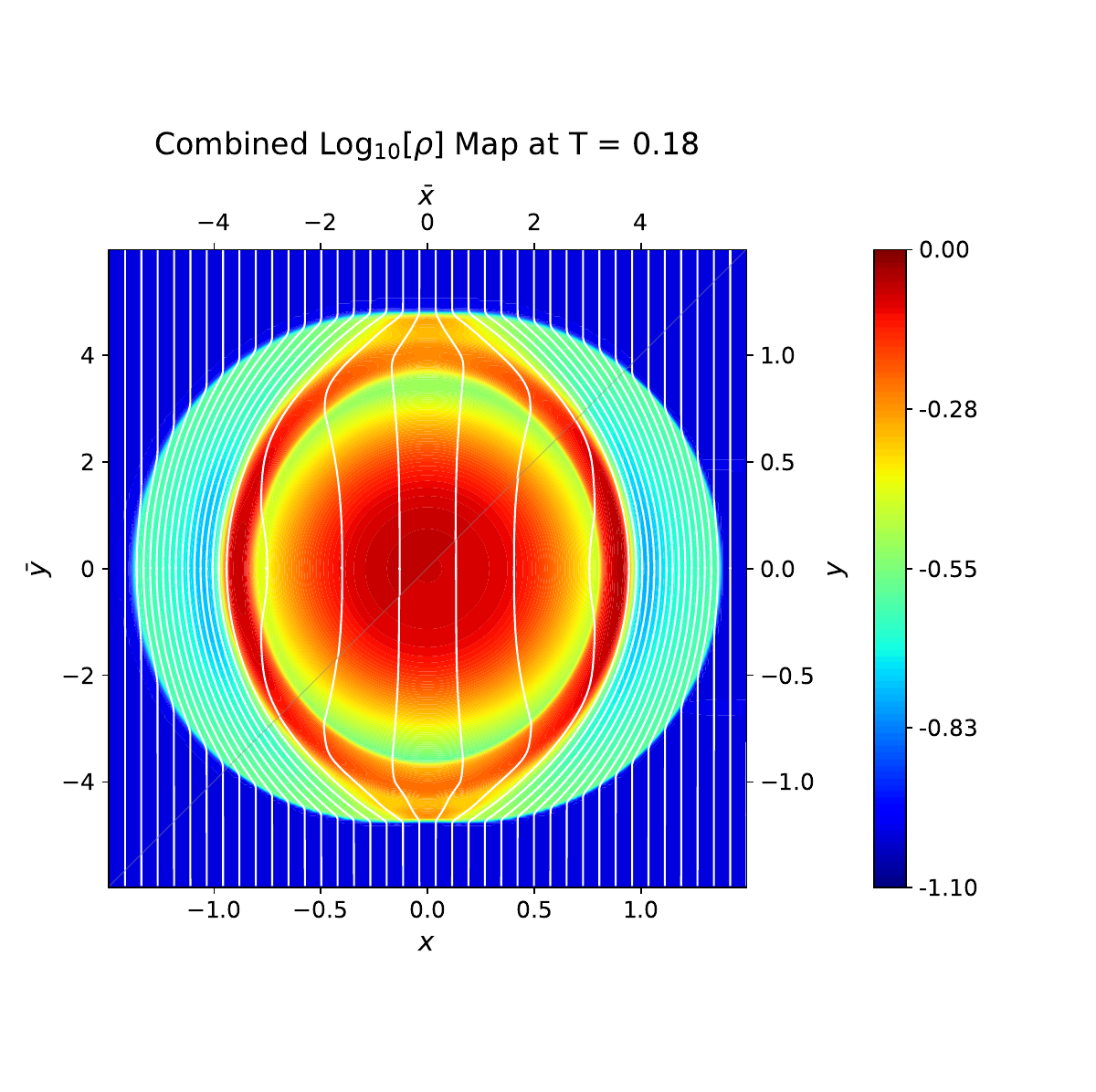}
        {\caption{2D Newtonian MHD blast wave problem. The lower-right half area (below the  counter-diagonal) shows the simulation results using the SSE coordinates, the upper-left half area (above the  counter-diagonal) shows the analogous outcome for the fixed coordinates. The color map represents the logarithm of the comoving density. White thin lines represent magnetic field lines. Simulations were performed on a $400 \times 400$ grid in both the fixed and SSE coordinates. }
       \label{fig:5}}
    \end{figure}

To check the performances of our approach in the Newtonian regime too, we repeated exactly the same 2D blast wave problem of the previous section, but adopting the Newtonian MHD limit in Eqs.~\ref{eq:newtini}-\ref{eq:newtfin}. The metric and the equation of state are left unchanged. However, due to the different dynamics, the final time is now $t=0.18$, for the same expansion factor of 4, so $\tau = t/4 = 0.045$. In Fig.~\ref{fig:5} we compare the results using the fixed and the SSE grid. As in the previous relativistic case, even now we see agreement in the location
and amplitudes of the various structures.  Possibly, there is some hint of a slightly sharper front along the $\bar{x} = x = 0$  direction.

\section{Conclusions}
\label{sec:conclusion}

In the present work, we proposed a simple method to introduce expanding (or contracting) grids within a conservative numerical scheme already designed for GRMHD astrophysical applications. We named this novel approach as “Self-Similar Expanding'' (SSE) coordinates method, and we validated it agaisnt standard special relativistic and Newtonian MHD tests.

Nowadays, many widely used packages for relativistic MHD are developed to be GRMHD friendly. We have shown here how, exploiting the advantages of a fully covariant $3+1$ formulation of the fluid equations, is it possible to manage expanding coordinate system in a native and easy way, retaining all the computational advantages of traditional Eulerian codes on fixed grid, either relativistic or Newtonian.  In this work we focused on analytical expansion laws, but this is not a limitation, because for many astrophysical systems the broad expansion properties can be derived from simple scaling arguments \citep{BMK76a, Sedov77, Weaver77a, chevalier82a}. Moreover our equations hold even if one define an expansion rate based on the on-the-fly result of the simulation run. 

In particular:
\begin{itemize}
    \item By exploiting the gauge invariance of the $3+1$ GRMHD formalism, our approach transforms standard solvers into expanding-grid codes with zero modifications to the core algorithm, requiring only a simple, one-line redefinition of the metric.
    \item The underlying SSE coordinate transformation leaves the Eulerian observer completely unchanged, naturally preserving simultaneity. Consequently, not only true Lorentz invariants, but all quantities defined with respect to the Eulerian observer, such as the Lorentz factor, individual kinetic terms (e.g., $v_{\bar{x}} v^{\bar{x}} = v_x v^x$), and magnetic field components (e.g., $B_{\bar{z}} B^{\bar{z}}=B_z B^z$), remain strictly invariant. This allows for a direct mapping of physical quantities between coordinate systems without the need for complex intermediate transformations.
    \item The proposed framework for SSE grids is highly extensible: non-ideal physical effects such as viscosity, heat conduction \citep{visc11,viscos13}, resistivity \citep{Komissarov07,Dionysopoulou15a}, and mean-field dynamos \citep{NBLDZ13}, as well as radiative cooling and heating, can be effortlessly incorporated by adopting standard covariant closures.
    \item Gravitational fields can be readily accommodated in the weak-field regime by setting the lapse function $\alpha = 1$ while maintaining $\partial_i\alpha \neq 0$ in the source terms of Eqs~\ref{eq:source_tot} and \ref{eq:source_sse}. One might even include self-gravity computed by solving Poisson equation on the expanding coordinates.
    \item Crucially, this method retains all the robust mathematical and computational properties of traditional fixed-grid Eulerian codes. Energy and momentum conservation are strictly enforced, and the solenoidal condition of the magnetic field is guaranteed to machine precision if using constrained transport methods rigorously based on the curl nature of the induction equation \citep{UCT04}. Furthermore, it is fully compatible with existing stretched-grid methodologies and can potentially be coupled with sophisticated AMR techniques operating directly on the expanding coordinates.
\end{itemize}

It should be explicitly noted that our methodological approach must be distinguished from frameworks based on transformations to a local comoving frame, such as standard cosmological simulations or fluid models intrinsically reliant upon local expanding or shearing boxes \citep{shearing, 1Grappin+93, Li+08a, ASTRONUM12, LDZ15, Toci2018, Echeverra+23a}. These latter techniques inherently entail a transformation to a system where a moving fluid appears stationary and, in the context of accelerating physical systems, necessitate the introduction of fictitious inertial forces \citep{Tenerani2017, Saldivia26}, a mathematical complication entirely absent within our SSE coordinate formulation. Furthermore, such conventional methodologies are fundamentally restricted to localized problem configurations, and it remains highly ambiguous whether, or by what rigorous mechanism, they might be robustly extended to encompass global computational domains.

Particularly interesting is the comparison with Arbitrary Lagrangian-Eulerian schemes (ALE), and in particular the recent General Relativistic implementation of \texttt{GIZMO} \citep{GIZMO25}. ALE schemes encompass a wide variety of method, from structured orthogonal grids \citep[e.g. \texttt{VH1},][]{VH1_12} to unstructured or meshless grids \citep[e.g. \texttt{BETHE-Hydro}, ][]{BETHE-HD12}. Because they can adapt to flows with both contracting and expanding features, they are more flexible than our globally expanding or contracting self-similar coordinates. However, implementing ALE in relativistic fluid dynamics introduces specific technical hurdles that our formulation avoids:
\begin{itemize}
    \item Flux Computation: most ALE schemes compute conservative fluxes by solving the Riemann problem on moving interfaces, which requires interpolating the internal 3-wave (or 7-wave in MHD) Riemann fan over the moving surface. Relativistic full-wave Riemann solvers are far more complex than their Newtonian counterparts \citep{Giacomazzo_Rezzolla06a}. Furthermore, this moving-interface approach is often incompatible with approximate solvers like our HLL, which efficiently provide fluxes by implicitly averaging over the Riemann fan.
    \item The Solenoidal Condition: maintaining divergence-free magnetic fields in ALE requires either divergence cleaning or solving the induction equation on moving edges. Divergence cleaning is known to underperform relative to Constrained Transport \citep{Tomida26a}, particularly in strongly shocked systems (ubiquitous in relativity, where the sound speed is always less than $c/\sqrt{3}$). Alternatively, the moving-edge approach necessitates complex interpolations of the electric field from the computational domain to the moving edges.
    \item Relativistic Formulations: deriving fluxes and electric fields on moving faces and edges in ALE requires explicit, full-fledged Lorentz transformations of the energy-momentum and Faraday tensors. In our native fully covariant formulation, grounded in General Relativity where motion is always relative to a coordinate system, these transformations are naturally taken care of by the fundamental form of the GRMHD equations.
\end{itemize}
Although we focused on analytically prescribed self-similar transformations for simplicity, our framework could in principle achieve ALE-like flexibility by adopting coordinate transformations that follow the mean flow as it evolves. For example, simply setting the shift vector equal to the mean flow speed directly yields the equations presented in the \texttt{GIZMO} paper cited above, which however retain several different source terms because they remain is the same coordinate system. While \texttt{ECHO} currently uses analytical metrics, dynamical General Relativistic codes routinely handle numerically evolving metrics, offering a potential pathway to full ALE functionality through a simple coordinate transformation.

By adopting a fully covariant, Eulerian $3+1$ approach to fluid dynamics and MHD, we illuminate the geometrical origin and clarify the mathematical properties of corrections terms that are typically used on expanding meshes, often derived with very close ties to specific numerical techniques (e.g. finite volumes) or even specific code implementations. We prove a general formalism to handle not just expanding coordinates, but any time dependence, including contracting, oscillating or non-isotropic cases.  This can be seen as a paradigm shift: instead of thinking in terms of a moving mesh with respect to a fixed physical coordinate system, the problem is faced in terms of a fixed mesh with respect to a dynamically moving coordinate system. The idea of decoupling the numerical mesh from the physical coordinates, encoding their relation through a metric tensor, can be applied not just to dynamical grids, but also to map physically stretched coordinates, to uniform numerical meshes.

In the present work we investigated the application of SSE grids just to special relativistic or Newtonian fluid dynamics and MHD, over flat space-times that are scale-free.  This notwithstanding, in principle, a transformation of coordinates in the spirit of the one proposed here could also be applied to any general relativistic metric, to investigate expanding (or contracting) astrophysical flows in the presence of strong gravity. However, if the foliation has singularities, like at the event horizon of the Schwarzschild metric in either radial-spherical or isotropic coordinates, one cannot employ global scale-free transformations (like our SSE), but needs to ensure that the transformation keeps the singularity out of the computational domain.

\begin{acknowledgements}
We acknowledge CINECA for the availability of HPC resources through a CINECA–INFN agreement (allocation INF26 - teongrav).  The authors thank the referee Ileyk El Mellah, for useful suggestions and in particular for pointing us to ALE schemes.
\end{acknowledgements}

\bibliographystyle{aa}
\bibliography{try}

\end{document}